\documentclass[aps,twocolumn,nofootinbib,preprintnumbers,prl,10pt]{revtex4-2}
\usepackage[utf8]{inputenc}
\usepackage[svgnames]{xcolor}
\usepackage[most]{tcolorbox}
\usepackage{caption}
\tcbset{colback=PaleGreen!0!white, colframe=NavyBlue, 
	highlight math style= {enhanced, 
		colframe=red,colback=red!10!white,boxsep=0pt}
}
\usepackage{relsize}

\usepackage{amsmath}
\usepackage[bottom]{footmisc}
\usepackage{braket}
\usepackage{graphicx}
\usepackage{mwe}
\usepackage[section]{placeins}
\usepackage{framed}
\usepackage{csquotes}
\usepackage{tikz}
\usetikzlibrary{decorations.markings}
\usetikzlibrary{decorations.pathmorphing}
\definecolor{shadecolor}{rgb}{0.90,0.90,0.90}
\usepackage{hyperref}
\usepackage{bbold}
\usepackage{subcaption}
\usepackage{pifont}
\usepackage{setspace}
\usepackage{amsmath, amssymb, amsthm, float, graphicx,amsfonts}
\usepackage{amsthm}
\usepackage{array, boldline, makecell, booktabs}

\theoremstyle{definition}

\hypersetup{colorlinks=true, linkcolor=DarkRed, citecolor=DarkRed,urlcolor=Indigo,linktocpage}
\def\beq{\begin{eqnarray}}\def\eeq{\end{eqnarray}}
\def\be{\begin{equation}}\def\ee{\end{equation}}
\def\bs{\begin{split}}\def\es{\end{split}}

\usepackage{bm}

\begin{document}

\title{\bf The String Dual to Free ${\cal N}=4$ Super Yang-Mills}
\author{\!\!\!\! Matthias R.~Gaberdiel$^{a}$\footnote{gaberdiel@itp.phys.ethz.ch} and Rajesh 
Gopakumar$^{b}$\footnote{rajesh.gopakumar@icts.res.in}\\ 
\it ${^a}$ Institut f\"ur Theoretische Physik, ETH Z\"urich \\
\hspace*{0.3cm} Wolfgang-Pauli-Stra{\ss}e 27, 8093 Z\"urich, Switzerland\\
\it ${^b}$ International Centre for Theoretical Sciences (ICTS-TIFR),\\
\it Shivakote, Hesaraghatta Hobli,
Bangalore North, India 560 089\\ }

\begin{abstract}{We propose a worldsheet description for the ${\rm AdS}_5\times {\rm S}^5$ string theory dual to large $N$, free ${\cal N}=4$ super Yang-Mills in four dimensions. The worldsheet theory is a natural generalisation of the recently investigated tensionless string on ${\rm AdS}_3\times {\rm S}^3\times \mathbb{T}^4$. As in the case of ${\rm AdS}_3$ it has a free field description, with spectrally flowed sectors, and is closely related to an (ambi-)twistor string theory. Here, however, we view it as a critical $N=4$ (closed) string background. We argue that the corresponding worldsheet gauge constraints reduce the degrees of freedom to a finite number of oscillators (string bits) in each spectrally flowed sector. Imposing a set of residual gauge constraints on this reduced oscillator Fock space then determines the physical spectrum of the string theory. Quite remarkably, we find compelling evidence that this prescription reproduces precisely the entire planar spectrum -- of single trace operators  -- of the free super Yang-Mills theory.}
\end{abstract}
\maketitle

{\noindent \bf Introduction}: 
The idea that gauge theories might be equivalent to string theories is around half a century old. However, it was only after the AdS/CFT correspondence that we have had examples of string theories describing a class of, large $N$, 4d (supersymmetric) gauge theories, as well as a dictionary between observables on both sides \cite{Maldacena:1997re, Gubser:1998bc, Witten:1998qj}. This remarkable connection between gravity and gauge theory has been the engine powering many advances in theoretical physics in the last couple of decades. 

One major limitation in these developments has been the intractability of the string theory side of the correspondence, beyond the large radius or supergravity limit. This is due to the presence of Ramond-Ramond flux in the background which is difficult to incorporate in the conventional RNS quantisation of strings. Thus, in the canonical example of type IIB string theory on ${\rm AdS}_5\times {\rm S}^5$ dual to 4d 
${\cal N}=4$ super Yang-Mills, most attempts to go beyond the supergravity limit have employed the Green-Schwarz formalism. But it has been difficult to quantise this theory even in a physical light cone gauge despite the presence of integrability (see \cite{Arutyunov:2009ga} for an overview). One exception has been the influential BMN limit \cite{Berenstein:2002jq} which takes a particular plane wave limit of the AdS background where the worldsheet theory becomes free. 
 
In this letter, we will propose a Green-Schwarz-like worldsheet description of string theory on ${\rm AdS}_5\times {\rm S}^5$, in the tensionless limit where it is believed to be dual to free ${\cal N}=4$ super Yang-Mills theory \cite{HaggiMani:2000ru}. Our proposal is motivated by our previous investigations \cite{Eberhardt:2018ouy, Dei:2020zui} into the tensionless limit of ${\rm AdS}_3\times {\rm S}^3\times \mathbb{T}^4$, whose equivalence with the dual free symmetric product 2d CFT has been established (at least at the level of correlators) in \cite{Eberhardt:2019ywk, Eberhardt:2020akk,Knighton:2020kuh}. Instead of a theory of two free fermions and bosons and their conjugates (all with conformal weight $\frac{1}{2}$ and first order kinetic terms), realising the ${\rm AdS}_3$ supergroup $\mathfrak{psu}(1,1|2)_1$ \cite{Eberhardt:2018ouy, Dei:2020zui}, we will now consider a theory of four such free fermions ($\psi^a$) and four bosons (${\lambda}^\alpha, \mu^{\dot{\alpha}}$), with their canonical conjugates, which realise $\mathfrak{psu}(2,2|4)_1$. As in the ${\rm AdS}_3$ case we will consider ``spectrally flowed" sectors of these fields,  see eqs.~\eqref{lambdaflow}-\eqref{ferflow2}. While we motivate this model and the approach to quantising it in the following sections (this will be further elaborated in a companion paper \cite{GG-comp}), here we focus on what we believe to be the upshot, which is easy to state in elementary terms. 

After gauge fixing, the physical degrees of freedom in the $w$-spectrally flowed sector, arise from the ``wedge modes" of half the bosons/fermions, namely 
\be\label{wedge}
 \mu^{\dot{\alpha}}_r\ , \  (\mu^\dagger_\alpha)_r\ ,  \  (\psi^\dagger_{1,2})_{r}  \ ,  \ \psi^{3,4}_{r} \ , \  \ \ (-\tfrac{w-1}{2} \leq r \leq \tfrac{w-1}{2}) \ .
\ee
These modes act nontrivially on the spectrally flowed vacuum state $|0\rangle_w$, and generate a Fock space of states \cite{foot0}. On this Fock space we then impose two
residual gauge constraints: 1) Every physical state must satisfy ${\cal C}_n|{\rm phys}\rangle =0$ for $n\geq 0$, where $[{\cal C}_n, \Phi_r]=\pm \frac{1}{2}\Phi_{n+r}$, and $\Phi_r$ stands for any of the above wedge modes; the modes of $\mu^\dagger_{\alpha}$ and $\psi^{\dagger}_{1,2}$ 
carry $(+)$ charge with respect to ${\cal C}_n$, while those of $\mu^{\dot{\alpha}}$ and $\psi^{3,4}$  carry $(-)$ charge. Also $\Phi_{m}\equiv 0$  if $m > \frac{w-1}{2}$, i.e.\ if the mode number lies outside the wedge; 2) the (generalised) mass-shell condition $L_0 =0 \,({\rm mod} \, w)$ on physical states, where $L_0$ counts, as usual, the mode number, i.e.\ $[L_n,\Phi_r] = (-\frac{n}{2} - r) \Phi_{n+r}$ \cite{foot0a}. 

Our central claim is this: {\emph{The resulting physical sector of the oscillator Fock space, for each $w$, is identical to the space of all the single trace operators built from $w$ super Yang-Mills fields and their derivatives.}} 

The oscillators, together with the above constraints capture the cyclicity of the trace, the null relations due to the free equations of motion, and give the right set of $\mathfrak{psu}(2,2|4)$ representations present in the free super Yang-Mills spectrum. A general proof for the agreement will be given in \cite{GG-comp}; we have also checked it explicitly at low levels, see below and \cite{GG-comp} for more details.

{\noindent \bf{The Worldsheet Theory}}:
The matter fields of our worldsheet theory comprise  the weight $\frac{1}{2}$ conjugate pairs of boson fields $(\lambda^\alpha, \mu^{\dagger}_{\alpha})$ and $(\mu^{\dot{\alpha}},\lambda^{\dagger}_{\dot{\alpha}})$, with $\alpha,\dot{\alpha} \in \{1,2\}$,  as well as four weight $\frac{1}{2}$ complex fermions $(\psi^a,\psi^\dagger_a)$, with $a\in\{1,2,3,4\}$. Here $\alpha$ and $\dot{\alpha}$ are spinor indices w.r.t.\ to two different $\mathfrak{su}(2)$'s, while $\psi^a$ transforms in the fundamental representation of $\mathfrak{su}(4)$. There is also a corresponding right-moving sector.  The modes of these fields obey the commutation relations 
\begin{align} \label{wsflds}
[\lambda^\alpha_r,(\mu^\dagger_\beta)_s] = & \delta^{\alpha}_{\beta} \, \delta_{r,-s} \ ,  \qquad
[\mu^{\dot{\alpha}}_r,(\lambda^\dagger_{\dot{\beta}})_s] = \delta^{\dot{\alpha}}_{\dot{\beta}} \, \delta_{r,-s} \ , \\
& \{\psi^a_r,(\psi^\dagger_b)_s \} = \delta^{a}_b \, \delta_{r,-s} \ ,
\end{align}
where (at least initially) $r,s\in \mathbb{Z} + \frac{1}{2}$. We can view these as components of ambitwistor fields $Y_{I}= (\mu^\dagger_{\alpha},\lambda^\dagger_{\dot{\alpha}},\psi^{\dagger}_a)$ and 
$Z^{I}=(\lambda^\alpha,\mu^{\dot{\alpha}},\psi^a)$, employing the notation of \cite{Berkovits:2004hg}. These matter fields have a vanishing net central charge, and the bilinears $Y_{I}Z^{J}$ generate the current algebra $\mathfrak{u}(2,2|4)_1$. The overall $\mathfrak{u}(1)$ generator ${\cal C} =\frac{1}{2} Y_IZ^I$ needs to be set to zero in order to restrict to $\mathfrak{psu}(2,2|4)$; it will  play an important role below. We note that each pair of modes $(r, -r)$ of the fields give rise to two copies of the usual oscillator construction of the $\mathfrak{psu}(2,2|4)$ superconformal algebra in 4d \cite{Gunaydin:1984fk}. Some relevant expressions are given in the appendix.  

In a sense, as in ${\rm AdS}_3$ \cite{Maldacena:2000hw}, all the nontrivial aspects of the worldsheet theory come from the spectrally flowed representations.  Let us define the $w$-spectrally flowed fields (with tildes) as \cite{foot0b}
\begin{align}
(\tilde{\lambda}^\alpha)_r = & (\lambda^\alpha)_{r-w/2}  \ ,  \quad 
(\tilde{\lambda}^\dagger_{\dot{\alpha}})_r = (\lambda^\dagger_{\dot{\alpha}})_{r-w/2} \ ,  \label{lambdaflow} \\ 
(\tilde{\mu}^{\dot{\alpha}})_r  =  & (\mu^{\dot{\alpha}})_{r+w/2}  \ , \quad (\tilde{\mu}^\dagger_\alpha)_r  =  (\mu^\dagger_\alpha)_{r+w/2}  \ , \label{muflow}  \\
(\tilde{\psi}^{a}_r) = & \psi^a_{r-w/2} \ , \quad 
(\tilde{\psi}^\dagger_{a})_r = (\psi^\dagger_{a})_{r+w/2}  \ \  (a=1,2) \ ,  \label{ferflow1} \\
(\tilde{\psi}^{b}_r) =  & \psi^b_{r+w/2} \ , \quad 
(\tilde{\psi}^\dagger_{b})_r =   (\psi^\dagger_{b})_{r-w/2}  
   \ \  (b=3,4) \ . \label{ferflow2} 
\end{align}
The precise form of the spectral flow is a natural generalisation of the one defined in \cite{Eberhardt:2018ouy, Dei:2020zui}. These modes act on the spectrally flowed vacuum state $|0\rangle_w$ which is characterised by the requirement that it is annihilated by all the tilde modes with positive (half-integer) mode number. In terms of the unflowed modes (without tildes) this means that we have a nontrivial action on this vacuum by the modes in eq.~(\ref{wedge}) with $r\leq \frac{w-1}{2}$. In addition, also the corresponding conjugate modes, i.e. 
\be\label{conjugate}
\lambda^\alpha_r\ , \ (\lambda^\dagger_{\dot{\alpha}})_r\ , \ (\psi^{1,2})_r \ , \  (\psi^{\dagger}_{3,4})_r  \ , 
\ee
with $r\leq - \tfrac{w+1}{2}$ do not annihilate $|0\rangle_w$ (and altogether they generate the full Fock space). Finally, we need to sum over all the $w$-flowed sectors with $w\in\mathbb{N}$.

The spectral flow also has a non-trivial effect on the dilatation operator ${\cal D}_0$ and the Cartan generators of $\mathfrak{su}(4)$, and we find explicitly 
\be\label{DRspecflow}
\widetilde{\cal D}_n = {\cal D}_n - w\, \delta_{n,0} \ , \qquad \widetilde{\cal R}_n = {\cal R}_n - w\, \delta_{n,0} \ , 
\ee 
where ${\cal R}_n= \frac{1}{2}[- ({\cal R}^1{}_1)_n - ({\cal R}^2{}_2)_n  +  ({\cal R}^3{}_3)_n +  ({\cal R}^4{}_4)_n]$, and $\widetilde{{\cal D}}_0 |0\rangle_w = \widetilde{{\cal R}}_0 |0\rangle_w =0$, see the appendix for details. It also shifts the worldsheet Virasoro generator as in the ${\rm AdS}_3$ case  
\be\label{tildeL}
\widetilde{L}_n = L_n - w({\cal D}_n - {\cal R}_n) \ .
\ee
Note that the combination $({\cal D}_0 - {\cal R}_0)$ is the BMN-like light cone Hamiltonian, which vanishes on the highest weight half-BPS state $|0\rangle_w$. It might seem that we are breaking the $\mathfrak{su}(4)$ invariance with our choice of spectral flow on the fermions eqs.~(\ref{ferflow1},\ref{ferflow2}), but this prescription only picks out a specific highest weight state, and the physical states lie in complete $\mathfrak{psu}(2,2|4)$ representations.    

Let us contrast our theory with the ${\rm AdS}_3\times {\rm S}^3$ case where we have half as many bosons and fermions, and a similar $\mathfrak{u}(1)$ generator, which has to be quotiented out in order to get the $\mathfrak{psu}(1,1|2)$ spacetime current algebra \cite{Eberhardt:2018ouy,Dei:2020zui}. (There is also a topologically twisted $\mathbb{T}^4$ sector in that case, which is absent for ${\rm AdS}_5\times {\rm S}^5$.) Inspired by \cite{Berkovits:1994wr,Berkovits:1999im} we propose that this (hybrid) Green-Schwarz worldsheet theory is quantised through its embedding in the $N=2$ string, with the above $\mathfrak{u}(1)$ constraint being implemented by the topologically twisted ${\nolinebreak N=2}$ algebra \cite{GGNVit}.  This then leads to two bosonic constraints (arising from the Virasoro and the $\mathfrak{u}(1)$ current of the $N=2$),  as well as the two fermionic ones (from the two supercurrents). Locally this removes four bosonic and four fermionic physical degrees of freedom. As seen in \cite{Eberhardt:2018ouy} we are left with an essentially  topological theory on ${\rm AdS}_3\times {\rm S}^3$, together with  the physical $\mathbb{T}^4$ excitations. We will comment more on this later.

We expect that we can quantise the present Green-Schwarz $\mathfrak{psu}(2,2|4)$ theory by an analogous embedding into the small $N=4$ string where the $\mathfrak{u}(1)$ constraint is again being implemented by the topologically twisted small $N=4$ algebra.  We note that the ghost system associated to this topologically twisted $N=4$ string has vanishing central charge so that the net central charge, after adding the matter piece, is still zero, see also \cite{Baulieu:1996mr}. The bosonic generators of the $N=4$ algebra consist of a  triplet of currents and the Virasoro generator, while we have now four fermionic supercurrents.
We therefore expect that imposing the $N=4$ constraints will locally remove eight bosonic and eight fermionic degrees of freedom, leaving only a topological sector behind. More specifically, we postulate that the $N=4$ conditions remove all $8$ bosonic and fermionic modes with $r\leq - \frac{w+1}{2}$, and we thus retain only the wedge modes of eq.~(\ref{wedge}). 

Let us pause to briefly explain the overall philosophy. The $N=2$ string embedding (of an $N=1$ theory in the RNS formalism), in the ${\rm AdS}_3\times {\rm S}^3$ case, is a way of making (some of) the spacetime SUSY manifest since this embedding allows for a map to Green-Schwarz like variables \cite{Berkovits:1993xq, Berkovits:1999im}. An embedding of the critical $N=2$ string in a small $N=4$ string has been given in \cite{Ohta:1995hw}. We believe such an extended gauge symmetry on the worldsheet is necessary for, and capable of, making the maximal spacetime SUSY of the ${\rm AdS}_5\times {\rm S}^5$ background manifest. The details of the embedding and the form of the resulting $N=4$ constraints are currently being worked out \cite{GGNVit}. Here we will simply assume that this can be carried out.

{\bf \noindent Matching the Spectrum:}
As we have argued above, in each $w$-spectrally flowed sector,  the $N=4$ constraints remove all but a finite number of modes, leaving behind only the wedge modes of eq.~(\ref{wedge}). 
%
%
The Fock space generated by these modes is the tensor product of $w$ copies of that generated by a single set of these oscillators. The latter generate what is called the singleton representation of $\mathfrak{psu}(2,2|4)$ once one imposes the condition of vanishing central charge ${\cal C}$ (see e.g.~\cite{Beisert:2004ry} and the appendix). The role of ${\cal C}$ in our sigma model is played by the $\mathfrak{u}(1)$ current discussed earlier. It is natural that physical states should be defined by requiring the residual constraints ${\cal C}_n|{\rm phys}\rangle =0$ (for $n=0,\ldots, w-1$). Similarly, we would expect that translation invariance along the discretised worldsheet made up by the $w$ string bits would be guaranteed by requiring the generalised mass-shell condition $L_0=0 \,({\rm mod} \, w)$. Since ${\cal C}_n |0\rangle_w = L_0 |0\rangle_w =0$ for $n\geq 0$, the spectrally flowed vacuum state $|0\rangle_w$ is a physical state by this criterion. We shall presently identify it with the half-BPS BMN vacuum.

In the following we would like to show how these physical states reproduce exactly the single trace spectrum of free super Yang-Mills theory. Let us define the generators 
\be
(S_{I}{}^{J}) _m = \sum_{r =  m - \frac{w-1}{2}}^{\frac{w-1}{2}} (Y_I)_{r} \, (Z^J)_{m-r} \ , 
\ee
where $Y_{I}$ and $Z^J$ were introduced before, see the paragraph below eq.~(\ref{wsflds}). 
For $m\geq 0$ these generators map our Fock space to itself, and they commute with ${\cal C}_n$ on this Fock space. The zero modes $(S_{I}{}^{J})_0$ map physical states to physical states and they furnish, in fact, a realisation of the global $\mathfrak{u}(2,2|4)$ symmetry; this shows, in particular, that the physical states fall into representations of the global $\mathfrak{psu}(2,2|4)$ algebra. For example, starting from the physical state $|0\rangle_w$ in the $w$-spectrally flowed sector, the $(S_{I}{}^{J})_0$ generate the full $\mathfrak{psu}(2,2|4)$ half-BPS multiplet with quantum numbers $(0,0;[0,w,0])_{w}$; here the first two entries are the 4d spins with respect to the two $\mathfrak{su}(2)$ subalgebras of $\mathfrak{psu}(2,2|4)$, while $[0,w,0]$ are the Dynkin labels of $\mathfrak{su}(4)$, and the subscript denotes the eigenvalue of ${\cal D}_0$ \cite{foot1}. (The non-trivial quantum numbers are a consequence of the shifts in eq.~(\ref{DRspecflow}).)  ln fact, we have checked for small values of $w$ that this BPS multiplet accounts for all the physical states in the sector with $L_0=0$. (For $w=2$ this also follows from our more general argument below.)

A relatively elementary counting argument suggests that there should be no physical states with $L_0>0$, and we have also verified this explicitly for some simple cases. This leaves us with the physical states with $L_0= - m w$ ($m>0$) which have, in general, a quite complicated structure. We have confirmed by explicit computations  (in particular we have checked all states with $w,{\cal D}_0\leq 4$) that they reproduce the intricate set of non-BPS multiplets (built from $w$ free fields) of the single trace spectrum of ${\cal N}=4$ SYM, enumerated, for example, in \cite{Bianchi:2003wx,Beisert:2003te,Beisert:2004di}. We will give more details of the comparison in our companion article \cite{GG-comp}. 

In the following we shall concentrate on the case $w=2$ where we can be more explicit. 
In this case the modes in eq.~(\ref{wedge}) have mode numbers $\pm \frac{1}{2}$. 
To describe the physical states with $L_0=-2m$, we observe that we can reduce any physical state by the application of suitable $(S_{I}{}^{J})_0$ generators to one that has the smallest 
number ($4m$) of oscillators, all with mode number $(+\frac{1}{2})$. (In particular, if we choose $I$ and $J$ to both correspond to the modes in (\ref{conjugate}), this reduces the number of modes by $2$.) Furthermore, all such states can be mapped into one another by the action of the $(S_{I}{}^{J})_0$ zero modes (where we now take one index to correspond to (\ref{conjugate}), and one to (\ref{wedge})). This shows that all the physical states at $L_0=-2m$ lie in an irreducible representation of $\mathfrak{psu}(2,2|4)$. 


We can actually identify these irreducible representations explicitly. For $L_0=0$ the physical states sit in the BPS multiplet $(0,0; [0,2,0])_{2}$ that is generated from $|0\rangle_2$ by the $(S_{I}{}^{J})_0$ modes. For $L_0=-2$, the physical state with smallest ${\cal D}_0$ eigenvalue is obtained by applying the $4$ fermionic $+\frac{1}{2}$ modes to $|0\rangle_2$; this leads to the highest weight state of the $\mathfrak{psu}(2,2|4)$ representation $(0,0;[0,0,0])_{2}$ -- the non-BPS Konishi multiplet. 
This construction generalises to $L_0=-2m$, for which the states with lowest ${\cal D}_0$ eigenvalue are 
\be\label{hspin}
\prod_{i=1}^{2m-2} \Bigl[ (\mu^\dagger_{\alpha_i})_{\frac{1}{2}} \, \mu^{\dot{\alpha}_i}_{\frac{1}{2}}  \Bigr] \,    (\psi^\dagger_1)_{\frac{1}{2}} (\psi^\dagger_2)_{\frac{1}{2}} \psi^3_{\frac{1}{2}} \psi^4_{\frac{1}{2}} \, |0\rangle_2  \ , 
\ee
and they generate the $\mathfrak{psu}(2,2|4)$ representation 
\be\label{L02m}
L_0 = - 2m : \qquad (m-1, m-1; [0,0,0])_{2m} 
\ee
since the indices $\{\alpha_i \}$ and $\{\dot{\alpha}_i \}$ are totally symmetrised. These are the higher spin conserved currents (bilinears) of the free theory with lowest component having spin $(2m-2)$. This then precisely reproduces the first line of \cite[eq.~(3.8)]{Beisert:2004di} -- the representation in eq.~(\ref{L02m}) is what is called ${\cal V}_{2m}$ there. These states, summed over $m$, combine into a single higher spin multiplet -- the symmetrised doubleton -- of the higher spin algebra $\mathfrak{hs}(2,2|4)$. 


We note that our oscillator construction is very reminiscent of the BMN states, except that here we are considering twistor variables. The wedge mode numbers are a discrete Fourier transform of the $w$ string bits each consisting of the singleton representation of $\mathfrak{psu}(2,2|4)$ \cite{bits}.  
Let us also note that for $w=0$ there are no physical states other than the NS vacuum.  
For $w=1$ the only physical states are in the $L_0=0$ sector (Ramond vacuum) and correspond to the singleton representation: a single copy of the super Yang-Mills fields, see the appendix for details. As usual, this is present only in a ${\rm U}(N)$ gauge theory and can be projected out. Thus the first nontrivial states arise from the above $w=2$ sector. 

{\bf \noindent Relation to the analysis for ${\rm AdS}_3$:} It is instructive to apply our approach to quantising the system to the case of  
${\rm AdS}_3\times {\rm S}^3\times \mathbb{T}^4$, i.e.\ to  consider the reduced oscillator Fock space for that case. The relevant wedge modes are then associated to $\mu^1$, $\mu_1^{\dagger}$,  $\psi^\dagger_1$, $\psi^3$, say, and we need to impose the constraints of ${\cal C}_n=0$, $L_0 =0 \,({\rm mod} \, w)$. Interestingly, we find a nice closed subsector of the states of the dual free symmetric product CFT. In some sense, these states are compactification independent and include, for $w=2$, the higher spin fields of $\mathfrak{hs}(1,1|2)$ \cite{Gaberdiel:2014cha}. The states for higher $w$ are special matter multiplets of the higher spin symmetry. We describe the details in \cite{GG-comp}.

{\bf \noindent Comments:} We noted earlier that the worldsheet fields can be viewed as ambitwistor string variables $Y_I$, $Z^I$ with the constraint ${\cal C}=\frac{1}{2}Y_IZ^I=0$ \cite{ambi}. There is also an analogue of spectral flow in that language since one needs to sum over instanton sectors carrying nontrivial ${\rm U}(1)$ flux. In fact,  in the presence of flux these sectors have a space of left-moving zero modes (the holomorphic sections of the nontrivial line bundle) for the twistors \cite{Dolan:2007vv, Nair:2007md}, which can be identified with our wedge modes. Viewing the wedge modes as (generalised) zero modes, motivates considering only the chiral (say left-moving) degrees of freedom, as we have implicitly done above; we should not expect the left- and right-moving wedge modes to describe independent degrees of freedom \cite{Amodel}. It would be interesting to understand the exact relation of our approach to quantising the theory as an $N=4$ string, to the (ambi)-twistor string approach to super Yang-Mills scattering amplitudes \cite{Witten:2003nn, Berkovits:2004hg, Mason:2013sva}. 


{\bf \noindent Discussion}: While we have invoked the superstructure of an embedding into a twisted $N=4$ string to quantise our worldsheet theory, it is clear that the reduced oscillator construction we have motivated, captures the physics of the free Yang-Mills spectrum. We expect correlators of our worldsheet theory to exhibit a similar localisation as in the 2d case \cite{Eberhardt:2019ywk, Dei:2020zui}. In fact, Feynman diagram contributions to free Yang-Mills correlators admit a geometric interpretation  in terms of covering maps to twistor space \cite{BGMR}, and generalise the correspondence with Strebel differentials \cite{Gopakumar:2005fx, Razamat:2008zr} already seen in \cite{Gaberdiel:2020ycd}. It would be interesting to also connect this to the picture of \cite{Berkovits:2019ulm}. The free Yang-Mills spectrum exhibits a Hagedorn transition \cite{Sundborg:1999ue, Polyakov:2001af, Aharony:2003sx}, and it would be nice to understand its physics in our worldsheet model. 

We note that once our proposed worldsheet theory is fully quantised, it should be possible to systematically consider perturbations away from the free point since there is a corresponding
marginal operator on the worldsheet. Given our string bit picture  we expect to  see a direct worldsheet reflection of the integrable spin chain Hamiltonians \cite{Beisert:2010jr}, and to be able to derive the AdS/CFT correspondence in a perturbative expansion.  

Finally, we must remark that our construction has a truncation, effectively, to the bosonic twistor oscillators ($\mu^{\dot{\alpha}}, \mu^\dagger_{{\alpha}})$ which then reproduces the planar spectrum of free, pure Yang-Mills. It is also intriguing that the twisted $N=4$ string embedding admits the $N=0$ (bosonic) string as a special vacuum, where the worldsheet SUSY is spontaneously broken \cite{Berkovits:1993xq}. Could this be the route to obtaining the long sought after string dual to planar QCD?

\section*{Acknowledgments} 
We thank Andrea Dei, Bob Knighton, Pronobesh Maity, Kiarash Naderi, and Vit Sriprachyakul for useful discussions and related collaborations. The research of MRG is supported by a personal grant of the Swiss National Science Foundation and more generally via the NCCR SwissMAP. R.G.'s research is partially supported by a J.C.~Bose fellowship of the DST as well as project RTI4001 of the Department of Atomic Energy, Government of India. 

\clearpage

\onecolumngrid
{\begin{center}\bf \Large{Supplementary material}\end{center} }

\section{Free Field Construction of $\mathfrak{psu}(2,2|4)_1$}
Here we give the free field realisation of the $\mathfrak{u}(2,2|4)_1$ generators. The generators for the bosonic subalgebra $\mathfrak{su}(2) \oplus \mathfrak{su}(2)\oplus \mathfrak{su}(4)$ are 
\begin{eqnarray}
{\cal L}^\alpha{}_\beta & = &  \mu^\dagger_\beta \, \lambda^\alpha - \tfrac{1}{2} \delta^\alpha_\beta  \, U\ ,  \\
\dot{\cal L}^{\dot{\alpha}}{}_{\dot{\beta}} & = &  \lambda^\dagger_{\dot{\beta}} \, \mu^{\dot{\alpha}} - \tfrac{1}{2} \delta^{\dot{\alpha}}_{\dot{\beta}}  \, \dot{U} \ , \\
{\cal R}^a{}_b & = & \psi^\dagger_b \psi^a - \tfrac{1}{4} \delta^a_b \, V \ , 
\end{eqnarray}
where normal ordering is understood, and we have introduced the $\mathfrak{u}(1)$ generators (mimicking what we did in \cite{Eberhardt:2018ouy})
\be
U  =  \mu^\dagger_\gamma \, \lambda^\gamma \ , \qquad 
\dot{U}  =  \lambda^\dagger_{\dot{\gamma}} \, \mu^{\dot{\gamma}} \ , \qquad 
V  =  \psi^\dagger_c\psi^c \ ,
\ee
which commute with ${\cal L}^\alpha{}_\beta$, $\dot{\cal L}^{\dot{\alpha}}{}_{\dot{\beta}}$ and ${\cal R}^a{}_b$. The modes of ${\cal L}^\alpha{}_\beta$ and $\dot{\cal L}^{\dot{\alpha}}{}_{\dot{\beta}}$ satisfy (separately) the commutation relations of the $\mathfrak{su}(2)_{-1}$, while the modes of ${\cal R}^a{}_b$ give rise to $\mathfrak{su}(4)_1$. The `off-diagonal' generators of $\mathfrak{u}(2,2|4)_1$ are 
\begin{equation}
\begin{array}{rclrcl}
{\cal Q}^a{}_\alpha & = & \psi^a\, \mu^\dagger_\alpha   \ , \qquad \qquad & {\cal S}^\alpha{}_a & = & \lambda^\alpha\, \psi^\dagger_a\,  \ , \\[4pt] 
\dot{\cal Q}^{\dot{\alpha}}{}_{a} & = & \mu^{\dot{\alpha}} \, \psi^\dagger_a \ , \qquad \qquad & \dot{\cal S}^a{}_{\dot{\alpha}} & = &  \psi^a \, \lambda^\dagger_{\dot{\alpha}} \ ,\\[4pt] 
{\cal P}^{\dot{\alpha}}{}_{\beta} & = &  \mu^{\dot{\alpha}}\, \mu^\dagger_\beta  \ , \qquad \qquad & {\cal K}^{\alpha}{}_{\dot{\beta}} & = & \lambda^\alpha \, \lambda^\dagger_{\dot{\beta}} \ ,
\end{array}
\end{equation}
and they satisfy (among others) the relations
\begin{eqnarray}
{}[({\cal K}^{\alpha}{}_{\dot{\beta}})_m,({\cal P}^{\dot{\gamma}}{}_{\delta})_n] & = & 
- \delta^{\dot{\gamma}}_{\dot{\beta}} \, ({\cal L}^{\alpha}{}_{\delta})_{m+n} + \delta^\alpha_\delta \, (\dot{\cal L}^{\dot{\gamma}}{}_{\dot{\beta}})_{m+n} 
- \delta^\alpha_\delta\, \delta^{\dot{\gamma}}_{\dot{\beta}} \, \bigl( {\cal D}_{m+n} + m \,\delta_{m,-n} \bigr) \ , \qquad \label{KP} \\ 
\{ ({\cal S}^{\alpha}{}_{a})_m,({\cal Q}^b{}_{\beta})_n \} & = & 
\delta^b_a \, ({\cal L}^{\alpha}{}_{\beta})_{m+n} + \delta^\alpha_\beta\, ({\cal R}^b{}_a)_{m+n} + \tfrac{1}{2} \delta^b_a\, \delta^\alpha_\beta\, \bigl({\cal D}_{m+n} + {\cal C}_{m+n} + 2m\, \delta_{m,-n} \bigr) \ ,  \\
\{ (\dot{\cal S}^{a}{}_{\dot{\alpha}})_m,(\dot{\cal Q}^{\dot{\beta}}{}_{b})_n \} & = & 
\delta^a_b \, (\dot{\cal L}^{\dot{\beta}}{}_{\dot{\alpha}})_{m+n} + \delta^{\dot{\beta}}_{\dot{\alpha}}\, ({\cal R}^a{}_b)_{m+n}  - \tfrac{1}{2} \delta^a_b\, \delta^{\dot{\alpha}}_{\dot{\beta}}\, \bigl ({\cal D}_{m+n} - {\cal C}_{m+n} + 2m\, \delta_{m,-n} \bigr) \ ,
\label{dotSdotQ}
\end{eqnarray}
where we have defined 
\be\label{BCD}
{\cal B}_n  =  \tfrac{1}{2} \, \bigl( U_n + \dot{U}_n \bigr) \ , \qquad 
{\cal C}_n  =   \tfrac{1}{2} \, \bigl( U_n + \dot{U}_n \bigr)  + \tfrac{1}{2} V_n  \ , \qquad 
{\cal D}_n  =  \tfrac{1}{2} \, \bigl( U_n - \dot{U}_n \bigr)   \ .
\ee
The generator ${\cal D}_n$ is the dilatation operator of $\mathfrak{psu}(2,2|4)_1$, and its zero mode will be identified with the conformal dimension in the 4d SYM theory. On the other hand, ${\cal B}_n$ and ${\cal C}_n$ play analogous roles to $Y_n$ and $Z_n$ in \cite{Eberhardt:2018ouy,Dei:2020zui}, respectively. In particular, ${\cal C}_n$ is central (except for the non-trivial commutator with ${\cal B}_n$), and needs to be `quotiented' in order to go to  $\mathfrak{psu}(2,2|4)_1$.

\textbf{\textit{Spectral Flow:}}
Under the spectral flow (\ref{lambdaflow})--(\ref{ferflow2}) the $\mathfrak{u}(1)$ currents transform as 
\be
\widetilde{U}_n = U_n - w \, \delta_{n,0} \ , \qquad 
\widetilde{\dot{U}}_n = \dot{U}_n + w \, \delta_{n,0} \ , \qquad 
\widetilde{V}_n = V_n \ , 
\ee
and hence $\widetilde{{\cal B}}_n = {\cal B}_n$ and $\widetilde{{\cal C}}_n = {\cal C}_n$, while ${\cal D}_n$ transforms as in eq.~(\ref{DRspecflow}). The diagonal generators of $\mathfrak{u}(4)$ are 
${\cal R}^{a}{}_a$, and they transform under spectral flow as 
\be\label{Rsf}
(\widetilde{{\cal R}}^{a}{}_a)_n  = \left\{ 
\begin{array}{cl}
({\cal R}^{a}{}_a)_{n} + \tfrac{w}{2} \delta_{n,0} \qquad & \hbox{if $a=1,2$} \\[4pt] 
({\cal R}^{a}{}_a)_{n} - \tfrac{w}{2} \delta_{n,0} \qquad & \hbox{if $a=3,4$,} 
\end{array}
\right.
\ee
while the off-diagonal generators ${\cal R}^{a}{}_b$ with $a\neq b$ transform as  
\be\label{Rsf1}
(\widetilde{{\cal R}}^{a}{}_b)_n  = \left\{
\begin{array}{cl}
({\cal R}^{a}{}_b)_{n+w}  \qquad & \hbox{if $b\in \{1,2\}$ and $a\in\{3,4\}$,} \\
({\cal R}^{a}{}_b)_{n-w}  \qquad & \hbox{if $a\in \{1,2\}$ and $b\in\{3,4\}$,} \\
({\cal R}^{a}{}_b)_{n}  \qquad & \hbox{if $a,b\in \{1,2\}$ or $a,b\in\{3,4\}$.} 
\end{array}
\right.
\ee
We take the generators of the form $({\cal R}^{a}{}_b)_0$ with $a<b$ to be the positive roots, and define the Cartan generators of $\mathfrak{su}(4)$ via
\begin{equation}
H_1  =  ({\cal R}^2{}_2)_0  -  ({\cal R}^1{}_1)_0\ , \qquad 
 H_2  =  ({\cal R}^3{}_3)_0  -  ({\cal R}^2{}_2)_0\ , \qquad
H_3  =  ({\cal R}^4{}_4)_0  -  ({\cal R}^3{}_3)_0 \ .
\end{equation}
Spectral flow then acts as 
\be\label{Hiflow}
(\widetilde{H}_1)_n = (H_1)_n \ , \qquad (\widetilde{H}_2)_n  = (H_2)_n - w\, \delta_{n,0}  \ , \qquad (\widetilde{H}_3)_n = (H_3)_n \ . 
\ee
Finally the generator ${\cal R}_n= \frac{1}{2}[-({\cal R}^1{}_1)_n - ({\cal R}^2{}_2)_n  +  ({\cal R}^3{}_3)_n +  ({\cal R}^4{}_4)_n]$ then transforms as (see eq.~(\ref{DRspecflow}))
\be
\widetilde{\cal R}_n = {\cal R}_n - w\, \delta_{n,0} \ . 
\ee

\section{The Singleton Representation}
It is illustrative to consider the $w=1$ case which gives rise to the basic building block of  super Yang-Mills, namely the singleton representation. In this case we only have the zero modes $\mu^{\dot{\alpha}}_0$, $(\mu^\dagger_\alpha)_0$,  $(\psi^\dagger_{1,2})_0$, $\psi^{3,4}_0$ acting on the Ramond vacuum state $|0\rangle_{w=1}$. We will drop the mode subscript in this section since there is no risk of confusion. It follows from (\ref{Rsf1}) and (\ref{Hiflow}) that $|0\rangle_{w=1}$ is a $\mathfrak{su}(4)$ highest weight state with Dynkin labels $[0,1,0]$. For the following it will be convenient to define the Ramond ground state via 
\be
|0\rangle_{\rm R} = \psi^\dagger_1\, \psi^\dagger_2\, |0\rangle_{w=1}\qquad \hbox{so that} \qquad 
\psi^\dagger_a |0\rangle_{\rm R} = 0 \quad (a=1,2,3,4) \ . 
\ee
[For $w=1$ it is convenient to consider the action of all fermionic zero modes on this state, not just of the wedge modes, where $\psi^{1,2} |0\rangle_{w=1} = \psi^\dagger_{3,4} |0\rangle_{w=1} =0$.] Since we have applied two ${\cal C}_0$ charged modes to $|0\rangle_{w=1}$, the ${\cal C}_0$ eigenvalue of $|0\rangle_{\rm R}$ is nonzero: ${\cal C}_0 \, |0\rangle_{\rm R} = |0\rangle_{\rm R}$. 

The physical state condition only requires imposing ${\cal C}_0\equiv {\cal C} =0$. It is then easy to verify that the physical states take the form 
\begin{align}
D^n\phi^{[ab]}\  & \longleftrightarrow  \ 
\prod_{i=1}^n(\mu^{\dot{\alpha_i}}\mu^\dagger_{\alpha_i}) \, \psi^a \psi^b \, |0\rangle_{\rm R} \ , 
 \\ 
D^n\Psi^{a \dot{\beta}} \  \longleftrightarrow \ 
\prod_{i=1}^n(\mu^{\dot{\alpha_i}}\mu^\dagger_{\alpha_i}) \, \mu^{\dot{\beta}} \psi^{a}\, \, |0\rangle_{\rm R} \ ,  \qquad 
& \qquad D^n\bar{\Psi}_{a {\beta}} \  \longleftrightarrow \ \epsilon_{abcd} 
\prod_{i=1}^n(\mu^{\dot{\alpha_i}}\mu^\dagger_{\alpha_i}) \mu^\dagger_{{\beta}}\, 
\psi^b\, \psi^c \, \psi^d\, |0\rangle_{\rm R}  \ , \\ 
D^nF^{\dot{\beta}\dot{\gamma}} \  \longleftrightarrow \ 
\prod_{i=1}^n(\mu^{\dot{\alpha_i}}\mu^\dagger_{\alpha_i}) \mu^{\dot{\beta}} \mu^{\dot{\gamma}} \, |0\rangle_{\rm R} \ , \qquad 
& \qquad D^nF_{{\beta}{\gamma}} \  \longleftrightarrow \  
\prod_{i=1}^n(\mu^{\dot{\alpha_i}}\mu^\dagger_{\alpha_i}) \mu^\dagger_{{\beta}}\mu^\dagger_{{\gamma}}\, \epsilon_{abcd} \psi^a\psi^b\psi^c\psi^d\, |0\rangle_{\rm R} \ ,
\end{align}
where we have also shown the correspondence to the on-shell states of the basic ${\cal N}=4$ super Yang-Mils multiplet (see e.g.\ \cite{Beisert:2004ry}). 
The $\phi^{[ab]}\equiv \phi^i$ are the six scalars in the fundamental of ${\rm SO}(6)$, i.e.\ the $[0,1,0]$ of $\mathfrak{su}(4)$; 
the ${\Psi}^{a\dot{\beta}}$  and $\bar{\Psi}_{a\beta}$ are the fermions, while $F_{\beta\gamma}$,  $F^{\dot{\beta}\dot{\gamma}}$ are the (anti)-self dual pieces of the gauge field strength. Finally $D^n$ is a shorthand for the $n$'th partial derivative, suppressing all Lorentz indices. These on-shell states comprise the singleton representation. 

All the single trace Yang-Mills operators (`words') are generated by taking the tensor product of $w$ copies of the singleton and projecting onto the cyclically invariant part. In a ${\rm U}(N)$ Yang-Mills theory, the singleton representation appears as the overall ${\rm U}(1)$ trace piece, and is therefore absent from the physical spectrum of an ${\rm SU}(N)$ gauge theory.


\begin{thebibliography}{99}
\bibitem{foot0}
Note that the $\dagger$ in $\mu^{\dagger}, \lambda^{\dagger}$ etc.\ does not refer to the complex or hermitian conjugate; these are independent oscillators.
\bibitem{foot0a} We also have ${\cal C}_n  |0\rangle_w = L_0 |0\rangle_w =0$ for $n>0$ since $\widetilde{L}_0\, |0\rangle_w =0$, see eq.~\eqref{tildeL}. As a consequence $|0\rangle_w$ is a physical state.
\bibitem{foot1}
In terms of the scalar field $\phi^i$ of super Yang-Mills, this is the  multiplet built from the completely symmetric traceless field ${\rm Tr}[\phi^{(i_1}\cdots \phi^{i_w)}]$ and its $\mathfrak{psu}(2,2|4)$ descendants.
\bibitem{Amodel}
A helpful parallel might be with the holomorphic A-model topological string which 
nevertheless captures the (non-chiral) de Rham cohomology of the target space.
\end{thebibliography}
\end{document}